\documentstyle[aps,pre,preprint,tighten,epsf,rotate]{revtex}
\squeezetable

\begin{document}

\preprint{TUM/T39-98-6}

\draft

\title{On the evaluation of some three-body variational integrals}
\author{Jos\'e Caro}
\address{Physik Department, Technische Universit\"at-M\"unchen,
D-85747-Garching, Germany}
\maketitle

\begin{abstract}
Stable recurrence relations are presented for the numerical computation
of the Calais-L\"owdin integrals
$$
\int \text{d} {\bf r}_1
\text{d} {\bf r}_2 \,
r_1^{l-1} r_2^{m-1} r_{12}^{n-1} \,
\exp{\{-\alpha r_1 -\beta r_2 -\gamma r_{12}\}}
$$
($l$, $m$ and $n$ integer, $\alpha$, $\beta$ and $\gamma$ real)
when the indices $l$, $m$ or $n$ are negative.
Useful formulas are given for particular values of the parameters
$\alpha$, $\beta$ and $\gamma$.
\end{abstract}


\section{Introduction}
When dealing with the three-body variational problem with Hylleraas basis,
it is usually necessary to make extensive use of integrals of the general form
\cite{CP62}
\begin{equation}
I(l,m,n;\, \alpha,\beta,\gamma) = \frac{1}{16 \pi^2}\int \text{d} {\bf r}_1
\text{d} {\bf r}_2 \,
r_1^{l-1} r_2^{m-1} r_{12}^{n-1} \,
\exp{\{-\alpha r_1 -\beta r_2 -\gamma r_{12}\}}\,,
\label{I}
\end{equation}
where $r_1= \left| {\bf r}_1 \right| $, $r_2=\left| {\bf r}_2 \right|$ and
$r_{12}=\left| {{\bf r}_2 - {\bf r}_1} \right|$.

For the case of $l$, $m$ and $n$ non-negative (that is, non-negative
powers of $r_1$, $r_2$ and $r_{12}$ once the volume element has been
taken into account), powerful, simple and stable recurrence relations that
permit the numerical calculation of these integrals can be found in
the literature \cite{SR67}. However, it is sometimes essential to have also an
expression for one of the integer indices being negative. For
instance, that happens in the atomic problem when one wants to
consider the mean value of the $r_{12}^{-2}$ operator \cite{TS77}
or relativistic corrections \cite{Br29}; or in the
nuclear problem when non-local terms are included in a Yukawa-like
interaction \cite{CG98}. In some cases, the integrals must be computed in
every step of the non-linear optimization procedure, and hence it is
clear the need of having a quick and reliable algorithm to compute
them. The specific cases $I(1,1,-1)$ and $I(0,-1,-1)$ were already
considered in Refs.\ \cite{TS77} and \cite{FH87} respectively.
For $\gamma = 0$ much work has been done
\cite{SR67,Ro65,SL66,Ha72,TJ76,Ki91},
also including explicitly the coupling of the angular momentum of the
two dynamical particles \cite{YD96}. Some work
has been devoted to the
analogous integrals for four- or more-body problems
\cite{FH87,Ro65,Ki91,PK94,YD97}.

The method proposed in this work to obtain the integrals\ (\ref{I})  is
specially useful when the same exponential
coefficients $\alpha$, $\beta$ and $\gamma$ appear in several
elements of the variational basis.

\section{General properties of
$\bbox{I(\lowercase{l},\lowercase{m},\lowercase{n})}$}
To study the general properties of the integral\ (\ref{I}) for $l$, $m$ and $n$
(possibly negative) integer numbers and $\alpha$,
$\beta$ and $\gamma$ real it is convenient to make use
of perimetric coordinates \cite{Pe58},
\begin{equation}
\left.
\begin{array}[c]{lcr}
u & = &  - r_1 + r_2 + r_{12}\\
v & = &    r_1 - r_2 + r_{12}\\
w & = &    r_1 + r_2 - r_{12}\\
\end{array}
\right\}
\label{perimetric}
\end{equation}
in terms of which the initial integral reads
\begin{equation}
I(l,m,n;\, \alpha, \beta, \gamma) = 2^{-(l+m+n+3)} \,
I_p\left(l,m,n;\frac{\beta + \gamma}{2},
\frac{\alpha + \gamma}{2},
\frac{\alpha + \beta}{2} \right)\,,
\end{equation}
where
\begin{equation}
I_p(l,m,n;\, a, b, c) = \int_0^\infty \text{d}u
\int_0^\infty \text{d}v
\int_0^\infty \text{d}w
\, (v+w)^l (u+w)^m (u+v)^n \,
\exp{\{-a u - b v - c w\}} \,.
\end{equation}
The integral $I_p$ is explicitly invariant under permutation of
conjugated pairs of parameters $\{ (l,a), (m,b), (n,c) \}$, and therefore
\begin{equation}
I(l,m,n;\, \alpha, \beta, \gamma) =
I(m,l,n;\, \beta, \alpha, \gamma) = 
I(n,m,l;\, \gamma, \beta, \alpha) \, ,
\label{genprosym}
\end{equation}
symmetry that will be used throughout this work.

The long range convergence of $I_p$ is ensured if $a$, $b$ and $c$ are
positive real numbers, that is, if
\begin{equation}
\alpha + \beta > 0\,, \quad
\alpha + \gamma > 0 \quad
\text{and} \quad
\beta + \gamma > 0 \,.
\label{long_range}
\end{equation}
That means that one of the exponentials parameters, $\alpha$, $\beta$ or
$\gamma$, can be zero or negative, provided that the other two are bigger
than the absolute value of the former. Note also that one of the exponential
coefficients of $I_p$ can be zero if the power of the
corresponding integration variable is negative and high enough. For instance,
$a=0$ with $l=0$ and $m=n=-1$ would yield a convergent result. Anyhow,
this is an almost useless case for the variational problem, because for
higher power integrals (that very likely should also be considered)
$a=0$ would lead to divergent quantities. From now on, we assume that
the requirements\ (\ref{long_range}) are fulfilled.

The study of the short range convergence can be straightforwardly done
case by case. Summarizing, for $l$, $m$ and $n$ integer,
and $\alpha$, $\beta$ and $\gamma$ real such that the
conditions\ (\ref{long_range})  are fulfilled,
the integral\ (\ref{I}) is convergent if and only if
\begin{equation}
l \ge -1\,, \quad m \ge -1\,, 
\quad n \ge -1 \quad
\text{and}
\quad l+m+n \ge -2\,. 
\label{genprolmn}
\end{equation} 

To have a procedure to generate the whole set of integrals\ (\ref{I}) one needs
relations for the cases $I(l,m,-1)$ and $I(l,-1,-1)$ where $l$ and $m$
are non-negative.

As soon as we have checked that the
integral we are looking for is convergent, integration over one
parameter can be applied to lower the conjugated power,
\begin{equation}
I(l,m,n;\, \alpha, \beta, \gamma) = \int_\gamma^\infty \text{d} c
\,
I(l,m,n+1; \, \alpha, \beta, c) \,.
\label{genpro1}
\end{equation}
On the other hand, derivation can always be used to increase indices,
\begin{equation}
(-\partial_\alpha)^{p} I(l,m,n; \alpha, \beta, \gamma) =
I(l+p,m,n; \alpha, \beta, \gamma)\,.
\label{genpro2}
\end{equation}
These properties, together with
\begin{equation}
I(0,0,0;\alpha,\beta,\gamma) =
(\alpha+\beta)^{-1} (\alpha+\gamma)^{-1} (\beta+\gamma)^{-1} \,,
\label{genpro3}
\end{equation}
are useful to derive all the integrals. Note also that for $\lambda > 0$
\begin{equation}
I(l,m,n;\lambda \alpha,\lambda \beta,\lambda \gamma) =
\lambda^{-(l+m+n+3)} I(l,m,n; \alpha,\beta,\gamma)\,,
\label{genprohom}
\end{equation} 
that is, for given $l$, $m$ and $n$,
$I$ is a homogeneous function of $\alpha$, $\beta$ and $\gamma$.
This fact, together with properties\ (\ref{genprosym}) and (\ref{genpro2})
yields a quite general recursion. Indeed,
differentiating with respect to $\lambda$ in the equation above
one gets the recurrence relation
\begin{eqnarray}
(l+m+n+3) I(l,m,n;\alpha,\beta,\gamma) & = &
\alpha I(l+1,m,n;\alpha,\beta,\gamma)
+ \beta  I(l,m+1,n;\alpha,\beta,\gamma)
\nonumber \\ & &{}
+ \gamma I(l,m,n+1;\alpha,\beta,\gamma)\,,
\label{genprorec}
\end{eqnarray}
valid for well defined integrals, in our case
$l$, $m$ and $n$ fulfilling conditions\ (\ref{genprolmn}).
In general, this recursion is of little utility, for to use it downwards,
which is the obvious direction,
one would have to know the value of the integrals on a plane
$l+m+n = \text{constant}$. We will take profit of a particular case
of Eq.\ (\ref{genprorec}) in section\ \ref{secmm}.

\section{Case $\bbox{I(\lowercase{l},\lowercase{m},-1)}$ with
$\bbox{\lowercase{l},\lowercase{m} \ge 0}$}

For the family of integrals $I(l,m,-1;\alpha,\beta,\gamma)$
with $l,m \ge 0$, a variation of the method exposed
in Ref.\cite{SR67} can be applied.  The recurrence relation that one  gets is
the following,
\begin{equation}
I(l,m,-1; \alpha, \beta, \gamma) =
\frac{1}{\alpha + \beta}
\left[ l I(l-1,m,-1) + m I(l,m-1,-1) + B(l,m) \right]\,,
\label{Im1rec}
\end{equation}
where
\begin{equation} B(l,m; \alpha, \beta, \gamma) = l!\, m!\,
\int_\gamma^\infty \text{d} c\, (\alpha+c)^{-l-1} \,
(\beta+c)^{-m-1}\,,
\label{Blmdef}
\end{equation}
which is a symmetric function under $(l,\alpha) \leftrightarrow (m,\beta)$
exchange, can be obtained through the relation
\begin{equation}
B(l,m) = \frac{1}{\alpha - \beta} \left[ l B(l-1,m) - m B(l,m-1) +
  C(l,m) \right] \,.
\label{Blmrecuns}
\end{equation}

Here the function $C(l,m)$  reads
\begin{equation} C(l,m;\alpha,\beta,\gamma) = \left\{
\begin{array}{ll}
(m-1)! \, (\beta + \gamma)^{-m} & \quad \text{if} \quad l=0
\quad \text{and}  \quad m > 0 \\
-(l-1)! \, (\alpha + \gamma)^{-l} & \quad \text{if} \quad l> 0
\quad \text{and}  \quad m = 0 \\
\log{(\alpha + \gamma) \over (\beta+\gamma)}& \quad \text{if} \quad l= 0
\quad \text{and} \quad m = 0 \\
0 & \quad \text{otherwise}
\end{array}
\right. 
\end{equation}
and is defined so that the recursion\ (\ref{Blmrecuns}) holds
also for $l=m=0$ although $B(0,-1)$  and $B(-1,0)$ are divergent. 
Note that $C(l,m)$ is antisymmetric under $(l,\alpha) \leftrightarrow
(m,\beta)$.

Unfortunately, in the  recursion\ (\ref{Blmrecuns}) subtractions
are involved, and hence one must look over the stability against
roundoff, in particular when $\alpha$ and $\beta$ are close to each
other.

It is also  possible to relate $B(l,m)$ to Gauss hypergeometric
function, ${}_2F_1$ \cite{AS72}, 
yielding
\begin{equation}
B(l,m; \alpha,\beta,\gamma) =
\frac{l! \, m!}{m+l+1} \, (\alpha+\gamma)^{-l-1} \,(\beta+\gamma)^{-m} 
\, {}_2F_1(1,l+1; m+l+2; z)\,,
\label{Blmhg}
\end{equation}
where $z\equiv\case{\alpha-\beta}{\alpha + \gamma}$. 
The use of the integral representation of the hypergeometric
function gives 
\begin{equation}
B(l,m; \alpha,\beta,\gamma) = (l+m)! 
\, (\alpha+\gamma)^{-l-1} \,(\beta+\gamma)^{-m} 
\,\int_0^1 {\rm d}t \, \frac{t^l \, (1-t)^m}{1-z t}\,.
\label{Blmint}
\end{equation}

\begin{figure}
\begin{center}
\leavevmode\rotate[r]{\hbox{
\epsfysize=8.6cm\epsfbox[56 41 564 712]{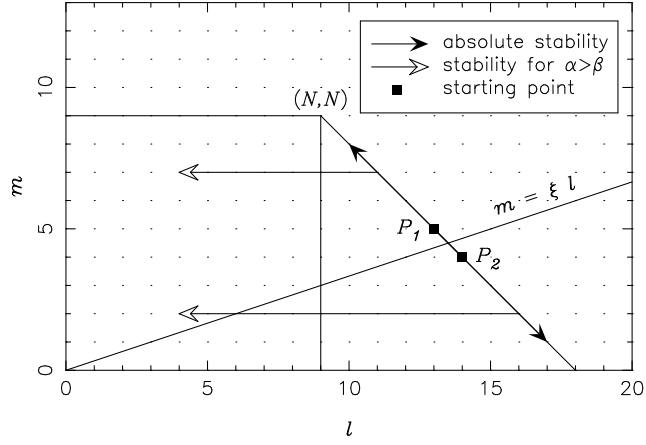}}}
\end{center}
\caption{Stability lines of the recursions for the calculation of
$I(l,m,-1)$. The solid arrows refer to the stable flux of
the recursion\ (\ref{Blmrecd}). The open  ones refer to
the recursion\ (\ref{Blmrecl}), but only if $\alpha>\beta$. In the opposite
case, the symmetric of Eq.\ (\ref{Blmrecl}) under
$(l,\alpha)\leftrightarrow (m,\beta)$ exchange can be used to move
downwards with stability.
}
\label{f-stab}
\end{figure}

From the definition\ (\ref{Blmdef}) it is possible to prove the equation
\begin{equation}
B(l+1,m) + B(l,m+1) = l!\, m!\, (\alpha + \gamma)^{-(l+1)}\,
(\beta + \gamma)^{-(m+1)} \qquad(l,m\ge 0)\,.
\end{equation}
Plugging this relation in Eq.\ (\ref{Blmrecuns}) yields
\begin{equation}
(l+m)\, B(l-1,m) - (\alpha - \beta)\, B(l,m) - (l-1)!\, m! \,
(\alpha + \gamma)^{-l}\,
(\beta + \gamma)^{-m} 
= 0\,,
\label{Blmrecl}
\end{equation}
valid for $m\ge0$ and $l>0$.
This equation permits to lower one unit the index $l$ of $B(l,m)$
with numerical stability if $\alpha > \beta$. In the opposite case,
the symmetry of $B(l,m)$ can be used to lower the index $m$
(see Fig.\ \ref{f-stab}).

On the other hand, using the Gauss relations for contiguous hypergeometric
functions one obtains 
\begin{equation}
m \,B(l+1,m-1) + (m-l\, \xi) \,B(l,m) - l \,\xi\, B(l-1,m+1) = 0\,,
\label{Blmrecd}
\end{equation}
where $\xi \equiv 1-z =\case{\beta + \gamma}{\alpha + \gamma}$.
This relation defines a recursion that can be
used to move on the diagonals $m+l=\, \text{constant}$.
As it is shown in Fig.\ \ref{f-stab}, the straight line 
$m(l) = \xi \, l$ on the $l$-$m$-plane separates the stability
regions of the recursion\ (\ref{Blmrecd}), so that
one can move with stability from this line in diagonal steps.

The final recipe to compute the set of $I(l,m,-1; \alpha, \beta, \gamma)$
for $l,m \le N$ is the following (see Fig.\ \ref{f-stab}). First,
two $B$'s are to be computed numerically to the required accuracy,
namely
\begin{equation}
B\left(\left[\case{2 N}{1+\xi}\right], 2 N -
\left[\case{2 N}{1+\xi}\right] \right)
\text{ and }
B\left(\left[\case{2 N}{1+\xi}\right]+1, 2 N -
\left[\case{2 N}{1+\xi}\right] - 1\right)
\label{2bs}
\end{equation}
(respectively, points $P_1$ and $P_2$ in Fig.\ \ref{f-stab}).
Then the recursion\ (\ref{Blmrecd}) is used to generate all needed starting
points to use  the recursion\ (\ref{Blmrecl}) leftwards (downwards)
if $\alpha > \beta$ ($\alpha < \beta$). Finally, the $B$'s obtained
in this way are introduced in  Eq.\ (\ref{Im1rec}).
To generate the two initial $B$'s one can compute
the integral in Eq.\ (\ref{Blmint}) by Gauss-Legendre quadrature. To optimize
the computation of the quadrature a change of variable is needed.
First, we use the symmetry of $B(l,m)$ to render $0 \le z < 1$. Next, we apply
in Eq.\ (\ref{Blmint}) the change of variable ($t\to s=s(t)$)
\begin{equation}
s(t) = \left\{
\begin{array}{cl}
t & \quad \text{if} \quad 0\phantom{.0} \le z < 0.8\\
\log{(2 - z - t)} &
\quad \text{if} \quad  0.8 \le z \le 0.99\\
\end{array}
\right.
\label{b-chov}
\end{equation}
For values of $z$ greater than 0.99 the hypergeometric function 
can be computed using the transformation formula 15.3.11 of the
Ref.\ \cite{AS72}. With the prescription above, more than fifteen
stable figures were obtained using 32 Gauss-Legendre points for
$2 N \le 60$. Note that the prescription\ (\ref{b-chov}) has been optimized
for the computation of the two initial $B$'s given in
the expression\ (\ref{2bs}), and
will not provide a similar accuracy for arbitrary values of $l$ and $m$.

The particular case $\alpha=\beta$ is specially simple. Indeed,
in that case the  $B$ function to be included in Eq.\ (\ref{Im1rec}) is
\begin{equation}
B(l,m; \alpha, \alpha, \gamma) = \frac{ l!  m!}{l+m+1}
(\alpha + \gamma)^{-(l+m+1)} \, ,
\end{equation}
and the calculations are numerically stable.
The case $\alpha=\beta$ is not only a mere academic example.  In many
practical problems the variational basis is chosen
so that any element has the same exponential coefficient both for
the coordinates $r_1$ and $r_2$. If the physical problem
requires to deal with $I(l,m,-1)$ integrals, then it is sensible to check
whether such a basis can produce the required accuracy.
This selection was successfully used in the context of a nuclear
theory problem \cite{CG98}.

In Table\ \ref{t-Im} we give some particular values
of $I(l,m,-1;\alpha,\beta,\gamma)$ with fourteen significant figures
to provide the reader with checking points.
\section{Case $\bbox{I(\lowercase{l},-1,-1)}$ with $\bbox{\lowercase{l}\ge0}$}
\label{secmm}
To generate the set of integrals $I(l,-1,-1)$ use can be made of
the relation
\begin{eqnarray}
\alpha\, I(l+1,-1,-1; \alpha,\beta,\gamma)  & = & 
(l+1) I(l,-1,-1;\alpha,\beta,\gamma)
\nonumber\\& & {}
- \beta I(l,0,-1;\alpha,\beta,\gamma)
- \gamma I(l,-1,0;\alpha,\beta,\gamma)\,,
\label{Im1m1rec}
\end{eqnarray}
which is valid for $l\ge0$. This equation is easily obtained
as a particular case of recurrence\ (\ref{genprorec}).
For $l=0$ direct calculation yields
\begin{equation}
I(0,-1,-1; \alpha,\beta,\gamma)  = \frac{1}{2 \alpha}
\left[
\frac{\pi^2}{6} - \log{\frac{\alpha+\gamma}{\beta+\gamma}}
\log{\frac{\alpha+\beta}{\beta+\gamma}}
-\text{Li}_2\left(\frac{\beta-\alpha}{\beta+\gamma}\right)
-\text{Li}_2\left(\frac{\gamma-\alpha}{\beta+\gamma}\right)
\right]
\label{Im1m1l0}
\end{equation}
(see also Ref.\ \cite{FH87}),
where $\text{Li}_2(z)=-\int_0^z \rm{d}y\,y^{-1} \log(1-y)$
is the dilogarithm function.

As said, the expression\ (\ref{genprorec}) is not applicable for
the case $l=m=n=-1$. Instead, one gets
\begin{equation}
\alpha I(0,-1,-1;\alpha,\beta,\gamma)
+ \beta  I(-1,0,-1;\alpha,\beta,\gamma)
+ \gamma I(-1,-1,0;\alpha,\beta,\gamma) = \frac{\pi^2}{4}\,.
\end{equation}
To fix the constant in the right hand side,
we had to make explicit use of the expression\ (\ref{Im1m1l0}).

The recursion\ (\ref{Im1m1rec}), which in general
is numerically unstable upwards,
can be used with stability to decrease the index $l$ if
$\alpha>0$, which is the interesting case in physics. But then,
one needs as starting point the integral with the highest wanted $l$.
As it can be derived from Eqs.\ (\ref{genpro1}-\ref{genpro3}), that integral
can be obtained through the computation of the quadrature
\begin{equation}
I(l,-1,-1;\alpha,\beta,\gamma) = \frac{l!}{2} \,
\left(\frac{2}{\beta+\gamma}\right)^{l+1} \,
\int_0^1 {\rm d}t \, \frac{1}{t} \,
G_l\left(\frac{2 \alpha}{\beta+\gamma},\frac{2 \beta}{\beta+\gamma};
\frac{1}{t}\right)\,,
\label{imm-in}
\end{equation}
where we have defined
\begin{eqnarray}
G_l(a,b;y) & = &\frac{1}{(a+y)^{l+1}} \left\{
\log\frac{(a + b + 2 (y-1)) (a - b + 2 y)}{(a + b) (a -b + 2)}
\right.
\phantom{aaaaaaaaaaaaaaaaaaaaaaaa} \nonumber\\ & & \hspace{-2cm} \left.
+ \sum_{m=1}^l \frac{1}{m} \left[
\left(\frac{a + y}{a + b}\right)^m  +
\left(\frac{a + y}{a - b + 2}\right)^m
- \left(\frac{a + y}{a - b + 2 y }\right)^m
- \left(\frac{a + y}{a + b + 2 (y -1) }\right)^m
\right]
\right\}\,.
\nonumber\\
\label{Gl}
\end{eqnarray}
Note that the integrand is positive, and that the sum in the
function $G_l$ is very efficiently computed in a single loop.
For values of $\alpha$, $\beta$ and $\gamma$ of the same order of magnitude
the quadrature converges very quickly for not very small values
of $l$ ($l>5$). This is not the case when one of the parameters is
larger than the other, but then a simple change of
variable helps to recover convergence. For instance, the following
prescription of changes of variable ($t\to s=s(t)$)
\begin{equation}
s(t) = \left\{
\begin{array}{cl}
\log{\left(t+\frac{\beta+\gamma}{2 \alpha}\right)} &
\quad \text{if} \quad  \alpha > \frac{l}{10} (\beta+\gamma)\\

\log{\left(1 + \frac{\alpha + \min\{\beta,\gamma\}}{2 (\beta+\gamma)}
- t \right)} 
& \quad \text{otherwise}
\end{array}
\right.
\label{imm-chov}
\end{equation}
produces the accuracies shown in Table\ \ref{t-acc-G}.
For $l=10$ (respectively, 20, 30  and 40) more than 3800 integrals per second
(respectively, 2500, 1800 and 1400) were  obtained  with more than
fourteen stable figures (32 Gauss-Legendre points, see Table\ \ref{t-acc-G})
in an inexpensive computer (a PC with a 200 MHz processor).

The prescription\ (\ref{imm-chov}) is not useful if both $\beta$ and
$\gamma$ are much smaller than $\alpha$ (e.g.,
$\beta, \gamma  < 0.01 \alpha$). However, the integrals for this case can
be safely generated without significant loss of accuracy
using the recursion\ (\ref{Im1m1rec}) upwards.
The relative error in the $l$-th integral accumulated because of
cancellations, which grows with $l$, can be then approximated as
\begin{equation}
{\cal E}(I(l,-1,-1)) =
\frac{\pi^2}{4} \, \frac{l!}{\alpha^{l+1}} \,
\frac{\text{machine-precision}}{I(l,-1,-1)}\,.
\end{equation}
For example, $I(60,-1,-1; 1,0.01,0.01) \simeq 0.768 \cdot 60!$,
and the relative error due to cancellations  in the repeatedly use
of\ (\ref{Im1m1rec}) to obtain this integral is  only of about three times
the machine-precision. For smaller values of $\beta$ and $\gamma$ the
accuracy is bigger. Note that Eq.\ (\ref{Im1m1l0}) is not appropriate
to evaluate $I(0,-1,-1)$ when $\beta$ and $\gamma$ are
smaller than $\alpha$. A better expression for this case is
\begin{eqnarray}
I(0,-1,-1; \alpha,\beta,\gamma)
& = &\frac{1}{4 \alpha}
\left[
\pi^2
- 2 \, \text{Li}_2\left(\frac{\beta+\gamma}{\alpha+\beta}\right)
- 2 \, \text{Li}_2\left(\frac{\beta+\gamma}{\alpha+\gamma}\right)
\right.\nonumber\\ & & \left.{}
- \log{\frac{\alpha+\beta}{\alpha}}
  \log{\frac{(\alpha+\beta)(\alpha+\gamma)}{(\alpha-\gamma)^2}}
- \log{\frac{\alpha+\gamma}{\alpha}}
  \log{\frac{(\alpha+\beta)(\alpha+\gamma)}{(\alpha-\beta)^2}}
\right.\nonumber\\ & & \left.{}
+ 2 \, \log{\frac{\beta+\gamma}{\alpha}}
\log{\frac{(\alpha+\beta)(\alpha+\gamma)}{(\alpha-\beta)(\alpha-\gamma)}}
\right]\,,
\label{Im1m1l0p}
\end{eqnarray}
where the dilogarithm function can be computed through the
expansion $\text{Li}_2(x) = \sum_{k=1}^\infty \case{x^k}{k^2}$.
For $x<0.02$, corresponding to $\beta,\gamma<0.01\alpha$, eight terms
in this expansion are enough to obtain sixteen stable figures.

A few particular cases of $I(l,-1,-1)$ are readily obtained from
the recursion\ (\ref{Im1m1rec}). Indeed, for $\alpha=0$ and $l \ge 0$
we have
\begin{equation}
I(l,-1,-1;0,\beta,\gamma) = \frac{1}{l+1}
\left[\beta I(l,0,-1;0,\beta,\gamma)
+ \gamma I(l,-1,0;0,\beta,\gamma)
\right]\,.
\end{equation}
The specific case $\alpha=\beta=\gamma$ reads
\begin{equation}
I(l,-1,-1;\alpha,\alpha,\alpha) =
\frac{l!}{(2\alpha)^{l+1}} \, S_l\,,
\end{equation}
where the coefficients
\begin{equation}
S_l = \sum_{m=0}^\infty \left(\sum_{i=0}^{l+m}\frac{1}{i+1}\right)
\frac{2^{-m}}{l+m+1}
\end{equation}
are to be computed only once. For $l\le100$ one does not need more
than 52 terms to achieve sixteen stable figures in $S_l$
without using any numerical procedure to accelerate convergence. The first
of these coefficients is $S_0=\case{\pi^2}{6}$.
Finally, for $I(l,-1,-1;\alpha,0,0)$ one has
\begin{equation}
I(l,-1,-1;\alpha,0,0) = \frac{\pi^2}{4}\, \frac{l!}{\alpha^{l+1}}\,.
\end{equation}

Some values of $I(l,-1,-1;\alpha,\beta,\gamma)$
are presented in Table\ \ref{t-Imm}.

\section{Summary}
Some recurrence relations to compute the integrals\ (\ref{I}) for all negative
integer parameters ($l$, $m$ and $n$) have been presented.
The stability of these recursions has been investigated, and algorithms
have been given to use them without loss of  accuracy due to cancellations.
The integrals
$I(l,m,-1;\alpha,\beta,\gamma)$ (where $l,m\ge0$) can be generated at low
computing cost. For the integrals $I(l,-1,-1;\alpha,\beta,\gamma)$
a quadrature involving $N+1$ terms is needed,
where $N$ is the highest  required $l$ and $\alpha$ is assumed to be positive.
Specially simple algorithms are given for
the cases $I(l,m,-1;\alpha,\alpha,\gamma)$, $I(l,-1,-1;0,\beta,\gamma)$,
$I(l,-1,-1;\alpha,\alpha,\alpha)$ and $I(l,-1,-1;\alpha,0,0)$. 

\section*{Acknowledgements}
The author gratefully thanks L.L. Salcedo for helpful comments on
a previous version of the manuscript, and E. Buend\'\i a for
some references.
This work was supported by the {\em Direcci\'on
General de Ense\~nanza Superior} (Spanish Education and Culture Ministry)
through a postdoctoral grant and the project PB95-1204.

\begin{table}
\begin{tabular}{c*{4}{c}}
\multicolumn{5}{c}{I($l,m,-1;1,\beta,\gamma)$}\\
$(\beta,\gamma)$
&   $l=10,\,  m=0$  &   $l=10,\,  m=5$  &   $l=20,\,  m=15$  &   
$l=40,\,  m=10$  \\\hline
(0.05,0.05)
& 2.5097803893512 ($+$07)& 1.1550294761619 ($+$14)& 5.8496691184166 ($+$45)&
 3.7263930880406 ($+$65)\\
(0.05,0.20)
& 1.0069028364788 ($+$07)& 4.9856655589152 ($+$12)& 1.1263701557234 ($+$41)&
 1.5245149068463 ($+$64)\\
(0.05,1.00)
& 2.1231968038912 ($+$06)& 6.1362890397889 ($+$11)& 1.8814741050023 ($+$39)&
 2.5445500764172 ($+$63)\\
(0.05,2.00)
& 1.0610620020573 ($+$06)& 3.0134288353911 ($+$11)& 9.0355811977946 ($+$38)&
 1.2652815721954 ($+$63)\\
(0.05,5.00)
& 4.2435098606331 ($+$05)& 1.1992798691243 ($+$11)& 3.5754444925655 ($+$38)&
 5.0533341605295 ($+$62)\\
(0.20,0.05)
& 5.9511548712149 ($+$06)& 1.4149826065364 ($+$12)& 7.4333291536966 ($+$39)&
 4.6331742767202 ($+$61)\\
(0.20,0.20)
& 2.5245340864963 ($+$06)& 3.8096117209107 ($+$11)& 1.5683171148516 ($+$38)&
 1.3617534581186 ($+$61)\\
(0.20,1.00)
& 4.9479973346792 ($+$05)& 7.1119098709184 ($+$10)& 1.5019778106939 ($+$37)&
 2.7821828122703 ($+$60)\\
(0.20,2.00)
& 2.4501414762596 ($+$05)& 3.5414728355891 ($+$10)& 7.3432360158513 ($+$36)&
 1.3921604053583 ($+$60)\\
(0.20,5.00)
& 9.7730721646852 ($+$04)& 1.4149098737738 ($+$10)& 2.9191066335088 ($+$36)&
 5.5698482446321 ($+$59)\\
(1.00,0.05)
& 2.3506936424270 ($+$05)& 2.1738100662595 ($+$08)& 1.8638785541149 ($+$30)&
 1.3896745377911 ($+$53)\\
(1.00,0.20)
& 6.8960382056483 ($+$04)& 9.7511110023298 ($+$07)& 7.0043082969132 ($+$29)&
 2.2092326515384 ($+$51)\\
(1.00,1.00)
& 2.6754225852273 ($+$03)& 2.0193624551498 ($+$07)& 1.4992836690734 ($+$29)&
 1.5383073655665 ($+$49)\\
(1.00,2.00)
& 9.7180274971942 ($+$02)& 1.0013416984929 ($+$07)& 7.5130229486046 ($+$28)&
 6.9573280539039 ($+$48)\\
(1.00,5.00)
& 3.5921654332679 ($+$02)& 3.9932069716804 ($+$06)& 3.0070041814429 ($+$28)&
 2.7138972320215 ($+$48)\\
(2.00,0.05)
& 5.6374109387576 ($+$04)& 1.4250979709052 ($+$06)& 9.1546004403515 ($+$24)&
 2.9503794635963 ($+$49)\\
(2.00,0.20)
& 1.5196492752737 ($+$04)& 4.7607751878871 ($+$05)& 1.4095247283656 ($+$24)&
 1.9130948733180 ($+$47)\\
(2.00,1.00)
& 1.5320840275499 ($+$02)& 3.8998899902794 ($+$04)& 7.9782256684551 ($+$22)&
 8.8675800592916 ($+$40)\\
(2.00,2.00)
& 1.6829755898961 ($+$01)& 1.6286558844196 ($+$04)& 3.5727727922307 ($+$22)&
 8.6310437705915 ($+$39)\\
(2.00,5.00)
& 4.3389324770986 ($+$00)& 6.1452543996272 ($+$03)& 1.3850914702513 ($+$22)&
 2.9003763096969 ($+$39)\\
(5.00,0.05)
& 8.9272368511676 ($+$03)& 2.0896049079214 ($+$03)& 1.2958485756621 ($+$18)&
 4.7703597474580 ($+$44)\\
(5.00,0.20)
& 2.3569834307325 ($+$03)& 5.6914967826940 ($+$02)& 1.0121437234633 ($+$17)&
 2.4003665222216 ($+$42)\\
(5.00,1.00)
& 1.5138128168377 ($+$01)& 6.1945864757004 ($+$00)& 2.8220667212483 ($+$13)&
 8.7853148834688 ($+$33)\\
(5.00,2.00)
& 3.2220328921320 ($-$01)& 5.5803922147916 ($-$01)& 1.0214473659965 ($+$12)&
 1.1612978333117 ($+$28)\\
(5.00,5.00)
& 3.4970794973642 ($-$03)& 1.0465458754289 ($-$01)& 2.1723867163698 ($+$11)&
 1.6228394644288 ($+$24)\\
\end{tabular}
\caption{Some values for the integrals $I(l,m,-1; 1,\beta,\gamma)$.
For instance, $ I(20,15,-1;1,0.2,5) =  2.9191066335088 \cdot 10^{36}$.
}
\label{t-Im}
\end{table}

\begin{table}
\begin{tabular}{d*{4}{d}}
  $l$ & $N_{\text{GL}}$=10 & $N_{\text{GL}}$=16 & $N_{\text{GL}}$=24
& $N_{\text{GL}}$=32\\
\hline
  5 &   5.9 &   8.9 &  11.1 &  12.6\\
  6 &   6.1 &   9.4 &  12.1 &  13.9\\
  7 &   5.6 &   9.6 &  12.9 &  14.9\\
  8 &   5.6 &   9.7 &  13.6 &  15.4\\
  9 &   5.9 &  10.3 &  14.1 &  15.3\\
 10 &   5.4 &   9.3 &  14.4 &  15.3\\
 15 &   5.1 &   8.7 &  14.0 &  15.2\\
 20 &   5.2 &   8.3 &  13.6 &  15.0\\
 25 &   4.7 &   8.2 &  12.9 &  14.9\\
 30 &   5.0 &   8.1 &  12.1 &  14.8\\
 35 &   4.7 &   7.5 &  11.4 &  14.8\\
 40 &   4.5 &   7.0 &  11.0 &  14.6\\
\end{tabular}
\caption{Number of stable figures ($-\log_{10} (\text{relative-error})$)
in $I(l,-1,-1;\alpha,\beta,\gamma)$
obtained using different numbers of Gauss-Legendre points ($N_{\text{GL}}$)
in the quadrature\ (\ref{imm-in}) with the prescription\ (\ref{imm-chov}).
The ratio $\case{\beta}{\alpha}$ moved in the range $[0.01, 10^4]$, and 
$\case{\gamma}{\alpha}$ in $[10^{-4}, 10^4]$.}
\label{t-acc-G}
\end{table}

\begin{table}
\begin{tabular}{c*{4}{c}}
\multicolumn{5}{c}{$I(l,-1,-1;1,\beta,\gamma)$}\\
$(\beta,\gamma)$
&   $l=10$  &   $l=20$  &   $l=30$  &   $l=40$  \\\hline
(\phantom{1}0.01,\phantom{1}0.01)
& 6.6191662489867 ($+$06)& 3.6251447231256 ($+$18)& 3.2983926561168 ($+$32)&
 8.5733728363337 ($+$47)\\
(\phantom{1}0.01,\phantom{1}0.20)
& 1.7540859652228 ($+$06)& 5.3112685642066 ($+$17)& 3.3942282862188 ($+$31)&
 6.9857465097091 ($+$46)\\
(\phantom{1}0.01,\phantom{1}0.50)
& 6.8609259644037 ($+$05)& 2.0155302235322 ($+$17)& 1.3180581225243 ($+$31)&
 2.7475204208958 ($+$46)\\
(\phantom{1}0.01,\phantom{1}1.00)
& 3.3243271983196 ($+$05)& 9.9937617391729 ($+$16)& 6.5672190824270 ($+$30)&
 1.3709460938605 ($+$46)\\
(\phantom{1}0.01,\phantom{1}2.00)
& 1.6469713752668 ($+$05)& 4.9876597990304 ($+$16)& 3.2808597542740 ($+$30)&
 6.8512909302258 ($+$45)\\
(\phantom{1}0.01,\phantom{1}5.00)
& 6.5729176667975 ($+$04)& 1.9940611043105 ($+$16)& 1.3120390818053 ($+$30)&
 2.7401331921302 ($+$45)\\
(\phantom{1}0.01,10.00)
& 3.2854418466598 ($+$04)& 9.9695938948872 ($+$15)& 6.5599782224617 ($+$29)&
 1.3700392619726 ($+$45)\\
(\phantom{1}0.20,\phantom{1}0.20)
& 4.5604766596349 ($+$05)& 3.5127689695513 ($+$16)& 4.9063595251091 ($+$29)&
 2.0412224928657 ($+$44)\\
(\phantom{1}0.20,\phantom{1}0.50)
& 1.4318580613654 ($+$05)& 7.2132201828707 ($+$15)& 8.1925006515660 ($+$28)&
 3.0132668197157 ($+$43)\\
(\phantom{1}0.20,\phantom{1}1.00)
& 6.1814890074087 ($+$04)& 3.2566277622792 ($+$15)& 3.8027988040087 ($+$28)&
 1.4138666122607 ($+$43)\\
(\phantom{1}0.20,\phantom{1}2.00)
& 2.9651034471161 ($+$04)& 1.5963963424310 ($+$15)& 1.8718211282572 ($+$28)&
 6.9707881275722 ($+$42)\\
(\phantom{1}0.20,\phantom{1}5.00)
& 1.1742458080974 ($+$04)& 6.3522447488220 ($+$14)& 7.4555535110623 ($+$27)&
 2.7776640095065 ($+$42)\\
(\phantom{1}0.20,10.00)
& 5.8633250910678 ($+$03)& 3.1737770991220 ($+$14)& 3.7255334615001 ($+$27)&
 1.3880777482000 ($+$42)\\
(\phantom{1}0.50,\phantom{1}0.50)
& 2.7602780782403 ($+$04)& 2.1429404549106 ($+$14)& 3.1095312824899 ($+$26)&
 1.3602027847527 ($+$40)\\
(\phantom{1}0.50,\phantom{1}1.00)
& 7.8596287095799 ($+$03)& 4.2261129342848 ($+$13)& 5.2077776493094 ($+$25)&
 2.0657542275279 ($+$39)\\
(\phantom{1}0.50,\phantom{1}2.00)
& 3.2906652367190 ($+$03)& 1.8845025819044 ($+$13)& 2.3675994785514 ($+$25)&
 9.4583791413986 ($+$38)\\
(\phantom{1}0.50,\phantom{1}5.00)
& 1.2667740672550 ($+$03)& 7.3501151392020 ($+$12)& 9.2602020503771 ($+$24)&
 3.7039195719454 ($+$38)\\
(\phantom{1}0.50,10.00)
& 6.3029915005903 ($+$02)& 3.6623979936588 ($+$12)& 4.6157947923024 ($+$24)&
 1.8465337659004 ($+$38)\\
(\phantom{1}1.00,\phantom{1}1.00)
& 9.2233071085062 ($+$02)& 3.9012855025163 ($+$11)& 3.1367380768282 ($+$22)&
 7.6500307975383 ($+$34)\\
(\phantom{1}1.00,\phantom{1}2.00)
& 2.1119403628357 ($+$02)& 6.5569071489222 ($+$10)& 4.6001747109760 ($+$21)&
 1.0314129070250 ($+$34)\\
(\phantom{1}1.00,\phantom{1}5.00)
& 7.2422587486309 ($+$01)& 2.3596051913831 ($+$10)& 1.6728442527147 ($+$21)&
 3.7666888585022 ($+$33)\\
(\phantom{1}1.00,10.00)
& 3.5623476884278 ($+$01)& 1.1649070967644 ($+$10)& 8.2662607382138 ($+$20)&
 1.8620745501178 ($+$33)\\
(\phantom{1}2.00,\phantom{1}2.00)
& 8.9219060942223 ($+$00)& 6.4071580456640 ($+$07)& 8.8443126974508 ($+$16)&
 3.7177770572299 ($+$27)\\
(\phantom{1}2.00,\phantom{1}5.00)
& 1.3397156572731 ($+$00)& 7.4723481641427 ($+$06)& 9.1598047862917 ($+$15)&
 3.5719240038509 ($+$26)\\
(\phantom{1}2.00,10.00)
& 6.2645017020798 ($-$01)& 3.5448536509329 ($+$06)& 4.3596533454989 ($+$15)&
 1.7025891413660 ($+$26)\\
(\phantom{1}5.00,\phantom{1}5.00)
& 3.7697615446040 ($-$03)& 2.6042296175688 ($+$01)& 3.4861019106182 ($+$07)&
 1.4250284687126 ($+$15)\\
(\phantom{1}5.00,10.00)
& 6.8653568463440 ($-$04)& 3.7115610730907 ($+$00)& 4.4359819939145 ($+$06)&
 1.6880938843748 ($+$14)\\
(10.00,10.00)
& 4.5242997355095 ($-$06)& 7.2459128782254 ($-$05)& 2.2554462512897 ($-$01)&
 2.1460242640172 ($+$04)\\
\end{tabular}
\caption{Some values for the integrals $I(l,-1,-1; 1,\beta,\gamma)$.
For instance, $ I(10,-1,-1;1,0.01,10) =  3.2854418466598 \cdot 10^4$.
}
\label{t-Imm}
\end{table}

\end{document}